\documentstyle[12pt]{article}
\begin{document}
\title{Light in metric space-time and its deflection by the screw
dislocation}
\author{ Miroslav Pardy\\[5mm]
Prague Asterix Laser System, PALS\\
Za Slovankou 3, 182 21 Prague 8 \\
Czech Republic\\[5mm]
and\\[5mm]
Department of Physical electronics\\ and\\
The Laboratory of the Plasma Physics \\
Masaryk University \\
Kotl\'{a}\v{r}sk\'{a} 2, 611 37 Brno, Czech Republic\\
e-mail:pamir@physics.muni.cz}
\date{\today}
\maketitle
\vspace{20mm}

\begin{abstract}
We explain the geometrical meaning of the metric of space-time. Then,
we derive the light deflection caused by the screw dislocation in
space-time.
The derivation is based on the idea that space-time is a medium
which can be deformed in such a way that the deformation of space-time
is equivalent to the existence of metric which is equivalent to gravity.
The existence of the screw dislocation in the cosmology is hypothetically
confirmed
by observation of light bursts which can be interpreted as the annihilation of
the giant screw dislocations with anti-dislocations.
The origin of gravitational bursts are analogical to the optical ones.
Hubble telescope
is able to detect only the optical bursts. The gravitational bursts then can
be detected by LIGO, VIRGO, GEO, TAMA and so on. The dislocation theory
of elementary particles is discussed.

\end{abstract}
\vspace{5mm}


\newpage
\section{Introduction}

\hspace{3ex}
\baselineskip 15pt

There is a possibility that during the big bang, supernova explosion,
gravitational collapse, collisions of the high-energy elementary particles
and so on, the dislocations in space-time are
created. In this article we derive the deflection of light caused by
the screw dislocation in space-time.

In order to derive such deflection of light,
it is necessary to explain the origin of metric in the Einstein
theory of gravity.

Einstein gives no explanation of the origin of the
metric, or, metrical tensor. He only introduces the Riemann geometry
as the basis for the general relativity \cite{keny}.
He "derived" the nonlinear
equations for the metrical tensor \cite{chandra} and never
explained what is microscopical origin of the metric of space-time.
Einstein supposed that it is adequate to
the metric that it follows from differential equations as their solutions.
However, the metric has an microscopical origin similarly to the
situation where the phenomenological thermodynamics
has also the microscopical and statistical origin.

Let us remember the different origins of metric. First, let us show
that metric is generated by the coordinate transformations. We
demonstrate it using the spherical transformations:
\begin{equation}
x = r\sin\theta\cos\varphi, \quad y = r\sin\theta\sin\varphi,\quad
z = r\cos\theta .
\label{1}
\end{equation}

The square of the infinitesimal element is as follows:
\begin{equation}
ds^{2} = dx^{2} + dy^{2} + dz^{2} = dr^{2} + r^{2}d\theta^{2} +
r^{2}\sin^{2}\theta \; d\varphi^{2} .
\label{2}
\end{equation}

We see that the nonzero components of the metrical tensor are
\begin{equation}
g_{11} =1, \quad g_{22} = r^{2}, \quad g_{33} = r^{2}\sin^{2}\theta .
\label{3}
\end{equation}

For $r = const$, it is $dr = 0$ and the element of the length is
the element of the 2-dimensional sphere in the 3-dimensional space. The resulting
metric is not the three dimensional one but only two dimensional
in the 3D space.
So, in order
to generate the 2D metric, it is necessary to use the 3D
transformations
in 3D space.
If we want the generate the metric on the 3D sphere, then it is necessary
to use the 4D transformations for $x, y, z, \xi$, where $\xi$ is the
extra-coordinate. So, the metric is generated by the curvilinear
transformations. Einstein suggested the possibility that metric can be
generated by gravitational field. He created the general theory of relativity
and gravitation. Henri Poincar{\'e} never accepted the metric generated by the
gravitational field.

The Riemann element $(ds)^{2}$ is defined as
\begin{equation}
ds^{2} = g_{\mu\nu}dx^{\mu}dx^{\nu}
\label{4}
\end{equation}
and it is composed from the four infinitesimal coordinate differentials.
It means if we
want to generate the metric of this four dimensional space-time
by the coordinate transformations,
then it necessary to use the coordinate transformations in 5D space-time.
Or, in other words, it is necessary to introduce the extra-dimension.
Einstein radically
refused the extra-dimensions and he pedagogically
explained the curvature of a space-time by introduction
the metric which depends on the temperature of the surface
\cite{einst}. Of course
such explanation of the origin of metric is not generally  accepted in
the textbooks and monographs  \cite{rind}. It was only pedagogical
explanation. Some mathematicians \cite{nat1},\cite{nat2}
tried to proof that our space
is  three-dimensional and they automatically excluded the extra-dimensions.
However such proofs are misleading because we know from the Bertrand
Russel philosophy of mathematics that the mathematical theorems
are not existential. In other words, mathematics cannot say nothing
on the existence of electron, proton, quarks, strings and so on, because
these things does not follow from the mathematical axiomatic system.
They are things of the external world and not of the world of
mathematics. At the same time pure mathematics cannot predict any physical constant,
because every physical constant is of dynamical origin.

Extra-dimensions can be introduced only by the definition and the existence of
them cannot be mathematically proved. We know that the three dimensional
space was confirmed by the most precise theory in the history of physics -
QED, and it means that the extra-dimensions were not confirmed. Also the Planck
law of the blackbody radiation in 4D space differs form the Planck law in 3D
space.
Similarly in genetics, the existence and the form of the molecule DNA can be
considered as a proof of the three dimensionality of space. The
formation of galaxies in the 3D space substantially differs from the
formation of galaxies in the four dimensional space.

We know that the extra-dimensions can be compactificated.
However, there is no physical law
which enables compactification. Compactification is only the mathematical
method of the string theory.

Einstein avoids the extra-dimensionality and compactification. He uses argumentation
\cite{einst} on  the existence
of the noneuclidean geometry using the 2D hot plane,
where the magnitude of a rule changes from
point to point being dependent on the temperature at a given
point. This method  was also used by Feynman \cite{feyn}.
Rindler does not use this method of argumentation \cite{rind}.

\section{Deformation origin of the space-time metric}

So, the question we ask, is,  what is the
microscopical origin of the metric of space-time.
We postulate that the origin of metric is the specific
deformation of space-time continuum. We take the idea from
the mechanics of continuum and we apply it to
the space-time medium. The similar approach can be found
in the Tartaglia article and his e-print \cite{tart}, where
space-time is considered as a deformable medium.

The mathematical description of the three dimensional deformation is
given for instance in \cite{land1}. The fundamental quantity is the
tensor of deformation expressed
by the relative displacements $u^{i}$ as follows:

\begin{equation}
u_{ik} = \left(\frac {\partial u_{i}}{\partial x^{k}} +
\frac {\partial u_{k}}{\partial x^{i}} + \frac {\partial u^{l}}{\partial
x^{i}} \frac {\partial u_{l}}{\partial x^{k}}\right);\quad i, k = 1, 2, 3.
\label{5}
\end{equation}

The last definition can be generalized to the four dimensional situation
by the following relation:

\begin{equation}
u_{\mu\nu} = \left(\frac {\partial u_{\mu}}{\partial x^{\nu}} +
\frac {\partial u_{\nu}}{\partial x^{\mu}} +
\frac {\partial u_{\alpha}}{\partial x^{\mu}}
\frac {\partial u^{\alpha}}{\partial x^{\nu}}\right); \quad
\mu, \nu = 0,1, 2,3,
\label{6}
\end{equation}
with $x^{0} = ct, x^{1} = x, x^{2} = y, x^{3} = z.$

 In order to establish the connection between metric $g_{\mu\nu}$ and
deformation expressed by the tensor of deformation, we write
for the metrical tensor $g_{\mu\nu}$ of the squared space-time element

\begin{equation}
ds^{2} = g_{\mu\nu}dx^{\mu}dx^{\nu},
\label{7}
\end{equation}
the following relation

\begin{equation}
g_{\mu\nu} = (\eta_{\mu\nu} + u_{\mu\nu}),
\label{8}
\end{equation}
where

\begin{equation}
\eta_{\mu\nu} =
\left(\begin{array}{cccc}
1 & 0 & 0 & 0\\
0 & -1 & 0 & 0\\
0 & 0 & -1 & 0\\
0 & 0 & 0 & -1\\
\end{array}\right).
\label{9}
\end{equation}

Instead of work with the metrical tensor $g_{\mu\nu}$, we can work with
the tensor of deformation $u_{\mu\nu}$ and we can consider the general theory
of relativity as the four-dimensional theory of some
real deformable medium as a corresponding form of the metrical theory.
First, let us test the deformation approach to the space-time in case of the
nonrelativistic limit.

\section{The nonrelativistic test}

The Lagrange function of a point particle with mass $m$ moving in
a potential $\varphi$ is given by the following formula \cite{land2}:

\begin{equation}
L = -mc^{2} + \frac {mv^{2}}{2} - m\varphi .
\label{10}
\end{equation}

Then, for a corresponding action we have

\begin{equation}
S = \int L dt = -mc \int dt \left(c - \frac {v^{2}}{2c} +
\frac {\varphi}{c} \right) ,
\label{11}
\end{equation}
which ought to be compared with $S = -mc\int ds$. Then,

\begin{equation}
ds = \left(c - \frac {v^{2}}{2c} +  \frac {\varphi}{c} \right) dt.
\label{12}
\end{equation}

With $ d{\bf x} = {\bf v}dt$ and neglecting higher derivative terms, we have

\begin{equation}
ds^{2} = (c^{2} + 2\varphi) dt^{2} - d{\bf x}^{2} =
\left(1 + \frac {2\varphi}{c^{2}}\right)c^{2}dt^{2} - d{\bf x}^{2}.
\label{13}
\end{equation}

The metric determined by this $ds^{2}$ can be be obviously related
to the $u_{\alpha}$ as follows:

\begin{equation}
g_{00} =  1 + 2\partial_{0}u_{0} +
\partial_{0}u^{\alpha}\partial_{0}u_{\alpha} =
1 + \frac {2\varphi}{c^{2}}.
\label{14}
\end{equation}

We can suppose that the time shift caused by the potential is
small and therefore we can neglect the nonlinear term in the last
equation. Then we have

\begin{equation}
g_{00} =  1 + 2\partial_{0}u_{0}  = 1 + \frac {2\varphi}{c^{2}}.
\label{15}
\end{equation}

The elementary consequence of the last equation is

\begin{equation}
\partial_{0}u_{0} =  \frac {\partial u_{0}}{\partial (ct)} =
\frac {\varphi}{c^{2}},
\label{16}
\end{equation}
or,

\begin{equation}
u_{0} = \frac {\varphi}{c}t + const.
\label{17}
\end{equation}

Using $u_{0} = g_{00}u^{0}$, or, $u^{0} = g^{-1}_{00}u_{0} =
\frac {\varphi}{c}t$, we get with $const. = 0$ and

\begin{equation}
u^{0} = ct' - ct,
\label{18}
\end{equation}
the following result

\begin{equation}
t'(\varphi) = t(0)\left(1 + \frac {\varphi}{c^{2}}\right),
\label{19}
\end{equation}
which is the Einstein formula relating time in the zero
gravitational field to time in the gravitational potential
$\varphi$. The time interval $t(0)$ measured remotely is so called the
coordinate time and $t(\varphi)$ is local proper time.
The remote observer measures time intervals to be deleted and
light to be red shifted.
The shift of light frequency corresponding to the gravitational
potential is, as follows \cite{land2}.

\begin{equation}
\omega = \omega_{0}\left(1 + \frac {\varphi}{c^{2}}\right).
\label{20}
\end{equation}

The precise measurement of the gravitational spectral shift was made by Pound
and Rebka in 1960. They predicted spectral shift $\Delta\nu/\nu =
2.46 \times 10^{-15}$ \cite{keny}.
The situation with the red shift is in fact the 
closed problem and no additional measurement is necessary.

While we have seen that the red shift follows
from our approach immediately,
without application of the Einstein equations,
it is evident that the metric
determined by the Einstein equations can be expressed by the tensor of
deformation. And vice versa, to the every tensor of
deformation the metrical tensor corresponds.

\section{The deflection of light by the screw dislocation}

The problem of the light deflection by the screw dislocation is the
problem of the recent years \cite{katan1}, \cite{katan2}, \cite{mora},
\cite{padu}, \cite{andr}, and so on. The
motivation  was the old problem of the deflection of light
by the gravitational field which according to Einstein causes the
curvature of space-time.

We know from the history of physics that the deflection of light by the
gravitational field of Sun was first calculated by Henri Cavendish in 1784
and it was never published. The first published calculation was almost
20 years later in 1981 by the Prussian
astronomer Johann Soldner. Einstein's calculation in 1911 was 0.83
seconds of arc. Cavednish and Soldner predicted a deflection 0.875 seconds of
arc. So, the prediction of Cavendish, Soldner \cite{brown} and Einstein in 1911 were
approximately half of the correct value which was derived in 1919 by Einstein.

Einstein in 1911 used the principle of equivalence for the determination of
the light deflection. As was shown by Ferraro \cite{fer} the Einstein application of this
principle was incorrect. The correct application was given only by Ferraro
in order to get the correct value. The deflection of
light by the topological defects as dislocations, disclinations  and so on
was to my knowledge never calculated by Einstein. In the recent time
such calculation was performed by \cite{katan1}, \cite{katan2}, \cite{mora}
\cite{padu}, \cite{andr} and so on.
Here we use the different and more simple method
and the definition of the screw dislocations which differs from the
above authors.

According to \cite{land1}, the screw deformation in the mechanics of continuum
was defined by the tensor of deformation
which is in the cylindrical coordinates as
\begin{equation}
u_{z\varphi} = \frac {b}{4\pi r},
\label{21}
\end{equation}
where $b$ is the $z$-component of the
Burgers vector. The Burgers vector of the screw dislocation
has components $b_{x} = b_{y} = 0, b_{z} = b$. The Burgers vector is for the specific
dislocation a constant geometrical parameter.

The postulation of the space-time as a medium enables to transfer the notions
of the theory of elasticity into the relativistic theory of space-time and
gravity. The considered transfer is of course the heuristical
operation, nevertheless the consequences are interesting.
To our knowledge, the problem, which we solve is new.

We know that the metric of the empty space-time is defined by the
coefficients in the relation:

\begin{equation}
ds^{2} = c^{2}dt^{2} - dr^{2} - r^{2} d\varphi^{2}  - dz^{2}.
\label{22}
\end{equation}

If the screw deformation is present in space-time, then the generalized metric
is of the form:
\begin{equation}
ds^{2} = c^{2}dt^{2} - dr^{2} - r^{2} d\varphi^{2} - 2u_{z\varphi}dzd\varphi
- dz^{2},
\label{23}
\end{equation}
or,

\begin{equation}
ds^{2} = c^{2}dt^{2} - dr^{2} - r^{2} d\varphi^{2} -
\frac {2b}{4\pi r}dzd\varphi - dz^{2}.
\label{24}
\end{equation}

The motion of light in the Riemann space-time is described by the equation
$ds = 0$. It means, that from the last equation the following
differential equation for photon follows:

\begin{equation}
0 = c^{2} -{\dot r}^{2} - r^{2} {\dot\varphi}^{2} -
\frac {b}{2\pi r}\dot z \dot\varphi - {\dot z}^{2}.
\label{25}
\end{equation}

Every parametric equations which obeys the last equation are equation of
motion of photon in the space-time with the screw dislocation.
Let us suppose that the motion of light is in the direction of the z-axis.
Or, we write approximately:

\begin{equation}
r \approx a;\quad {\dot z} = v.
\label{26}
\end{equation}
Then, we get equation of $\varphi$:

\begin{equation}
2 \pi a^{3} \dot\varphi^{2} + b v\dot\varphi = 2\pi a(c^{2} - v^{2}).
\label{27}
\end{equation}
We suppose that the solution of the last equation is of the form

\begin{equation}
\varphi = At.
\label{28}
\end{equation}
Then, we get for the constant $A$  the quadratic equation

\begin{equation}
2\pi a^{3}A^{2} + bvA + 2\pi a(v^{2} - c^{2}) = 0
\label{29}
\end{equation}
with the solution

\begin{equation}
A_{1/2} = \frac {-bv \pm \sqrt{b^{2}v^{2} - 16\pi^{2} a^{4}(v^{2} - c^{2})}}{4\pi a^{3}}.
\label{30}
\end{equation}

Using approximation $v \approx c$, we get that first root is approximately
zero and for the second root we get:

\begin{equation}
A \approx \frac {-bc}{2\pi a^{3}},
\label{31}
\end{equation}
which gives the function $\varphi$ in the form:

\begin{equation}
\varphi \approx \frac {-bc}{2\pi a^{3}}t.
\label{32}
\end{equation}

Then,  if $z_{2} - z_{1} = l$ is a distance between two points on the straight
line parallel with the axis of screw dislocation then,
$\Delta t = l/c$, $c$ being the velocity of light.
For the deflection angle $\Delta\varphi$, we get:

\begin{equation}
\Delta\varphi \approx \frac {-bl}{2\pi a^{3}}.
\label{33}
\end{equation}

So, we can say, that if we define the screw
dislocation by the metric of eq. (24), then, the deflection
angle of light caused by such dislocation is given by eq. (33). The
result (33) is only approximative and we do not know what is the
accuracy of such approximation. This problem can be solved using
the approximation theory.

Let us remark that the exact trajectory of photon in the field of the screw
dislocation can be determined from the trajectory equation

$$\frac {d^{2}x_{\mu}}{ds^{2}} + \Gamma_{\mu}^{\alpha\beta}
\frac {dx_{\alpha}}{ds}\frac {dx_{\beta}}{ds} = 0. \eqno(34)$$
which was used in many textbooks. However, According to Landau et
al. [10], the equation is
contradictory for photon, because in this case  $ds = 0$, and it means
that the last equation is not rigorously defined. Landau et al. derived the
deflection of light from the Hamilton-Jacobi equation for particle with the
rest mass $m = 0$, which moves with the light velocity. However, this
approach is not also absolutely correct because in
the classical field theory it is
not possible to define photon. Photon is a quantum object. Rigorous
derivation of the deflection of light was given by Fok \cite{fok},
who used the
mathematical object "the front of wave" and his result is valid without any
doubt.

Let us remark that equation (34) has two meanings: geometrical and
physical. The geometrical meaning uses $g_{\mu\nu}$ which follows from
the curvilinear transformations and the physical meaning of
$g_{\mu\nu}$ is metric of the gravitational field calculated by means of the
Einstein equation. The second meaning is the Einstein
postulate and cannot be derived from so called pure mathematics.
Only experiment can verify the physical meaning of equation (34).

The problem of interaction of light with the gravitational field is not
exhausted by our example. We can define more difficult problems such
as deflection of the coherent light, laser light, squeezed light,
soliton light,
massive light with massive photons, light of the entangled photons and
so on. No of these problems was still solved because they are only for
brilliant experts very well educated. And this is the pedagogical problem.

\section{The physical generation of the screw dislocation of the space-time}

Now the question arises, how to determine the mechanical or electrodynamical
or laser system which will generate the screw dislocation in space-time.
We know, that for real crystals the generation of the screw dislocation is
the elementary problem of the physics of crystals. If we use Einstein's
equations, then the problem is mathematical one. Or, to see it,
let us write the Einstein equations with
the cosmological constant $\lambda$.

$$R_{\mu\nu} -\frac {1}{2}g_{\mu\nu}R + \lambda g_{\mu\nu} =
-\kappa T_{\mu\nu},
\eqno(35)$$
where $T_{\mu\nu}$ is the tensor of the energy and momentum.

In case of the perfect fluid and pressure, tensor of the energy and momemntum
is as follows:

$$T_{\mu\nu}({\rm mech}) = (\varrho + p)u_{\mu}u_{\nu} + pg_{\mu\nu},
\eqno(36)$$
where $\varrho$ is a density and $p$ is a pressure of the fluid. The
quantities $u_{\mu}$ are four velocities of the fluid.

In case that the tensor of the energy and momentum is created
electromagnetically, then,

$$T_{\mu\nu}({\rm elmag}) = \frac {1}{4\pi}
\left(F_{\mu\alpha}F^{\alpha}_{\nu} -
\frac {g_{\mu\nu}}{4}F^{\alpha\beta}F_{\alpha\beta}\right).\eqno(37)$$
where $F_{\alpha\beta}$ is the tensor of the electromagnetic field.

So, because $g_{\mu\nu}$ is determined as the metrical tensor corresponding
to the screw dislocation of the space-time, the left side of the Einstein
equations is given and the problem is to find the mechanical and
electrodynamical quantities, which determine corresponding tensors of energy
and momentum.

The surprising thing is the fact that if there is
no curvature of space time,
than thanks to the existence of the cosmological constant, the solution
corresponding to the mechanical or electrodynamical
systems is not absolutely zero.
The cosmological constant is in such a way very important quantity and
evidently cannot be zero. The dislocations of space-time are in harmony with the
cosmological constant.

Let us remark that Einstein equations were derived intuitively by
Einstein \cite{chandra} and rigorously by Hilbert
from the Lagrangian using the variational method \cite{keny}. The Hilbert derivation is
pure mathematical one and it means it is very simple.
This variational method enables to start relativity physics
from the general theory and to derive the special
relativity as a classical limit of the general theory. This approach was
presented by Rindler \cite{rind2}.
To our knowledge, such unconventional but very elegant approach
was not presented in any textbook on relativity theory.

The tensor of the energy momentum in equation (36) is
rigirously defined.
The problem is, how to identify the distribution of the cosmological
objects with this tensor. To our knowledge, there is no
mathematical theorem for such rigorous identification.

\section{Cosmological consequences}

The verification of our theory and at the same time the existence of the
nonzero cosmological constant can be performed during the measurement of the
cosmical microwave background radiation. In case of the existence of some
defects in space-time the distributions of this radiation will be
inhomogeneous and it will depend on the density and orientations of the
dislocations, screw dislocations, disclinations and other topological
defects
of the space-time.

It is evident that also in case that the curvature caused by
the some topological
defect is zero, then, thanks to the existence of the cosmological
constant in the Einstein equations the defects can be generated  mechanically
or electrodynamically as it follows from the Einstein
equations. So, investigation
of the cosmical microwave background can inform us on the distribution
of the topological deffects in the space-time and on the possible origins
of these deffects \cite{lachi}.

\section{The laboratory verification of a theory}

We can consider the situation which is analogical
to the space-time situation. In other words, we can consider 
the  modified Planckian experiment with the black body radiation.
The difference from the original Planck situation is that we consider
inside of the black body  some optical medium
with dislocations. Then, in case that the optical properties depend also on
the presence of dislocations, in other words, that the local index of refraction
depend on presence of dislocations,
then we can expect the modification of the
Planck law of the blackbody radiation.  We know that the most simple problem
with the constant index of refraction was calculated \cite{kubo}.
To our knowledge, although
this is only so called table experiment, it was never performed in some
optical laboratory. It is possible to expect that during the experiments
some  surprises will appear.
However, the practical situation can be realized, if we prepare some crystal
with the screw dislocations with given orientation.
Then, in case that the optical
properties are expressed by means of metric in crystal, the metric
will determine the optical path of light in the crystal and it means that the
screw dislocations can be investigated by optical methods and not only
by the  electron microscope.

\section{Discussion}

We have defined in the harmony with the author article  \cite{pard1},
\cite{pard2} gravitation as a
deformation of a medium called space-time. We have used
equation which relates Riemann metrical tensor to the
tensor of deformation of the space-time medium and applied
it to the gravitating
system, which we call screw dislocation in space-time. The
term screw dislocation was used as an analogue with
the situation in the continuous mechanics. We derived the angle of
deflection of light passing along the screw dislocation axis at the distance
$a$ from it on the assumption that trajectory length was $l$.
This problem was not considered
for instance in the Will monograph \cite{will}.The screw dislocation
was still not observed in space-time and it is not clear what role play
the dislocations in the development of universe after big bang.
Our method can be applied
to the other types of dislocations in space-time and there is no problem
to solve the problem in general. We have used here the
specific situation because of its simplicity.
We have seen that the problem of dislocation in space-time is
interesting
and it means there is some scientific value of this problem.

It is well known from the quantum field theory and experiment,
that every particle has its partner in the form of
the antiparticle \cite{mai}. For instance the
antiparticle to the electron is positron. It is well known that  after
annihilation of particle-antiparticle pair, photons are generated.
For instance

$$e^{+} + e^{-} \quad\rightarrow \quad 2\gamma .\eqno(38)$$

Now, if we define that to every dislocation $D$ exists antidislocation  $\bar D$,
(which can be considered also as an analogue of the antistring, \cite{sriva}),
then, we can write the following equation which is analogical to (38)

$$D + \bar D\quad \rightarrow\quad N\gamma\eqno(39)$$
where N is natural number  and gamma denotes photon. The next equation is
also possible:

$$D + \bar D\quad \rightarrow\quad Ng\eqno(40)$$
where $g$ denotes graviton.

It is possible also to consider the high-energy process with the incident
particles $a$ and $b$ as follows:

$$a + b \quad \rightarrow \quad c_{1} + c_{2} + c_{3} + ...   c_{n},
\eqno(41)$$
where $c_{i}$ are denotations of identical or different
particles. It is well-known that the equation  (41)
will be fundamental equation of LHC.

In case of the existence of the dislocations in universe, equations (39)
and (40)  represents the burst of photons or gravitons in the cosmical space.
So, we defined the  further possible interpretation of the photonic and
gravitational bursts in cosmical space.

Let us remark that the dislocation approach to the particle physics
are in  harmony with the Einstein dream and later
Misner-Wheeler geometrodynamics where all existing elementary
objects can be defined
as some form of space-time. Misner and Wheeler  \cite{mis}, \cite{whe}
consider also that neutrino is the
specific form of the space-time. Let us still remark that we know from
the history of philosophy that long time before Christ,  Anaximandros
introduced {\bf apeiron} as a medium from which all particles, and
therefore all visible universe was created.
So, we can say that the famous trinity of men, Einstein-Misner-Wheeler, is the follower of
Anaximandros.

The identification of the fundamental particles by the dislocations
is in harmony with the
relation for the energy of the dislocation  \cite{cot}

$$E = \frac{E_{0}}
{\sqrt{1 - \frac{v^{2}}{c^{2}}}},\eqno(42)$$
where $E_{0}$ is the energy of dislocation when its velocity $v$ is zero and
$c$ is the velocity of sound in the crystal. In case of medium called space-time
the velocity $c$ is the velocity of light in vacuum. So,  vacuum is in a certain sense
medium which is similar to the crystal medium. The analogy with the dislocations in crystal
is of course heuristical step which is the integral part of the
methodology of discoveries and it cannot be
rigorously algorithmically defined.

Although equation (42) can be identified with the relativistic equation for
dependence of
energy on velocity of elementary particle, we cannot identify
electron with the screw dislocation in space-time. Why? Because we know that
the attractive or repulsive force between two screw dislocations are different
than the force between two electrons, two positrons and electron and positron.
The second reason is that
the anomalous magnetic moment of electron is of the dynamical origin as it
follows from the Feynman diagram technique while
the classical dislocation does not involve such dynamics.
On the other hand, we do not know how other dislocations such as
circular dislocations, cylindrical dislocations, helix,  double
helix, triple helix dislocations and so on are related to
elementary particles, specially to neutrinos. We know that all
physical constant in the standard model are of the dynamical
origin, but at present time it
is not clear what is their derivation from the more fundamental
theory (subquark theory, string theory and so on),
or, from the dislocation theory of elementary particles.
We think that dislocation theory of elementary particles is not at
present time prepared to
give the answer to these difficult questions.

The 3D screw dislocation can be extended mathematically to the
N-dimensional space, or, space-time. However the interpretation of
the N-dimensional theory needs introducing of the compactification.

So, we can say that the Einstein dream of the unification of all
objects and forces in nature in the framework of geometrodynamics
is far from the successful realization because the identification of
ultimate blocks of nature with dislocations and with the topological defects
is not possible at this time.

According to Veltman, there are plenty of mysteries
in particle physics \cite{velt}.
We also know one mystery in mathematics. This is the imaginary
number $i = \sqrt{-1}$. It is considered usualy as the mysterious number
because there is no geometrical
meaning of this number. The mathematical relation

$$i^{i} = e^{-\pi/2}\eqno(43)$$
is not mysterious, because the proof of this relation is elementary.
In physics the situation is a such, that if we do not know what is the physical
meaning of some relation, then it is mysterious.

We know that Parmenides was not able to understand motion. Motion was
mysterious for him. Why? Because he was not able to introduce time in his
system of thinking.
In particle physics and in the string theory \cite{ant} the confinement of
quarks is mysterious and the solution of his problem in the dislocation theory
of fundamental constituents is open.
We think, that appropriate understanding of the stringlike
dislocations \cite{lund} and
definition of the ultimate building blocks of nature
can solve all problems of particle physics together with removing all mysteries.

While the verification of the existence of the optical bursts caused by the annihilations
of the giant screw dislocations and anti-dislocations can be detected
by the Hubble telescope,
the gravitational bursts can be probably
detected by LIGO \cite{ligo}, VIRGO \cite{virgo},
GEO \cite{geo}, TAMA \cite{tama}, and so on.

\end{document}